\documentclass[twocolumn,showpacs,preprintnumbers,amsmath,aps]{revtex4}
\usepackage{graphicx}
\usepackage{dcolumn}
\usepackage{bm}
\usepackage{color}
\begin{document}
\newcommand{\kvec}{\mbox{{\scriptsize {\bf k}}}}
\def\eq#1{(\ref{#1})}
\def\fig#1{Fig.\hspace{1mm}\ref{#1}}
\def\tab#1{table\hspace{1mm}\ref{#1}}
\title{
---------------------------------------------------------------------------------------------------------------\\
The superconducting phase of Calcium under the pressure at 200 GPa: the strong-coupling description}
\author{R. Szcz{\c{e}}{\`s}niak, A.P. Durajski}
\affiliation{Institute of Physics, Cz{\c{e}}stochowa University of Technology, Al. Armii Krajowej 19, 42-200 Cz{\c{e}}stochowa, Poland}
\email{adurajski@wip.pcz.pl}

\date{\today}
 
\begin{abstract}
The thermodynamic parameters of the superconducting state in Calcium under the pressure at $200$ GPa have been determined. 
The numerical analysis by using the Eliashberg equations in the mixed representation has been conducted. It has been stated, that the critical temperature ($T_{C}$) decreases from $36.15$ K to $20.79$ K dependently on the assumed value of the Coulomb pseudopotential ($\mu^{*}\in\left<0.1,0.3\right>$). Next, the order parameter near the temperature of zero Kelvin ($\Delta\left(0\right)$) has been obtained. It has been proven, that the dimensionless ratio $2\Delta\left(0\right)/k_{B}T_{C}$ decreases from $4.25$ to $3.90$ together with the growth of $\mu^{*}$. Finally, the ratio of the electron effective mass to the electron bare mass ($m^{*}_{e}/m_{e}$) has been calculated. It has been shown, that $m^{*}_{e}/m_{e}$ takes the high value in the whole range of the superconducting phase's existence, and its maximum is equal to $2.23$ for $T=T_{C}$.
\end{abstract}
\maketitle
%
\section{INTRODUCTION}

In Calcium under the influence of the high pressure ($p$), the whole sequence of the structural phase transitions is induced. 

In the normal conditions Calcium crystallizes in the fcc structure (Ca-I). For the pressure at $20$ GPa, the transition to the structure bcc (Ca-II) has been observed. The next change appears for the pressure's value at $32$ GPa, above which the structure sc (Ca-III) is realized \cite{Olijnyk}. In $2005$ Yabuuchi {\it et al.} have found another two crystal phases: Ca-IV and Ca-V \cite{Yabuuchi1}, \cite{Nakamoto}. It has been stated, that the phase Ca-IV is stable from $119$ GPa; above the pressure of $143$ GPa appears Ca-V. The phases Ca-IV and Ca-V have been indentified experimentally by Fujihisa {\it et al.} \cite{Fujihisa}. The following assignment has been proposed: Ca-IV with the $P4_{1}2_{1}2$ structure and Ca-V with the {\it Cmca} structure. In the range of pressures from $158$ GPa to $207$ GPa, Nakamoto {\it et al.} have discovered the phase Ca-VI (the {\it Pnma} structure) \cite{Nakamoto2}. Above, Sakata {\it et al.} have reported existence of the host-guest structure (Ca-VII) \cite{Sakata}.

The results achieved by using the {\it ab initio} methods only partially agree with the experimental scheme of the structural phase transitions. 
In particular, Ishikawa {\it et al.} and Arapan {\it et al.} have reproduced the appearance of the low pressure structures (fcc and bcc) \cite{Ishikawa}, \cite{Arapan}. However, the values of the pressure, at which the structural transitions appear, differ from the experimental values. Above the pressure at $36$ GPa, the sequences of the successive structural transitions, explicitly differ between themselves and do not reproduce the experimental scheme. In the context of the essential theoretical results, we have mentioned the paper of Yao {\it et al.} \cite{Yao}. The authors were the ones who identified the phase Ca-IV as the structure {\it Pnma} and the phase Ca-V as {\it Cmca}. 

In the range of the very high pressures (from $135$ GPa to $495$ GPa), the calculations of Ishikawa {\it et al.} suggest the stability of the structure $I4/mcm(00\gamma)$. However, the similar value of the enthalpy has the $\it Pnma$ structure (at last for the lower pressures from the discussed range) \cite{Ishikawa}. It has to be underlined, that the stability of the {\it Pnma} structure from $158$ GPa to $180$ GPa is suggested by the paper of Aftabuzzaman and Islam \cite{Aftabuzzaman}. Also, the results presented by Yin {\it et al.} prove especially low values of the {\it Pnma} enthalpy ($p\in\left<135, 220\right>$ GPa) \cite{Yin}. 

Besides the very complex sequence of the structural transitions, which to the present day has not been finally determined, Calcium is being characterized with the interesting superconducting properties. In particular, the very high values of the critical temperature among the simple metals. 

The possibility of the superconducting state's existence in Calcium has been reported in $1981$ by Dunn and Bundy \cite{Dunn}. In $1996$, the definite experiment has been performed by Okada {\it et al.} \cite{Okada}. After ten years, the studies of Okada have been repeated by Yabuuchi {\it et al.} \cite{Yabuuchi2}. The researchers determined the dependence of the critical temperature on the pressure, at the same time ascertaining, that $T_{C}$ appreciably exceeds the values determined by Okada. 

According to Yabuuchi {\it et al.}, the superconducting state in Calcium is induced in the following crystal phases: Ca-III, Ca-IV and Ca-V \cite{Yabuuchi2}. In particular, in the phase Ca-III the critical temperature quickly increases together with the pressure's growth (from about $3$ K for $p=58$ GPa to $23$ K for $p=113$ GPa). In the phase Ca-IV, the critical temperature is the subject of the saturation, while in the phase Ca-V it takes the value of $25$ K at $161$ GPa. In 2011, Sakata {\it et al.} have reported $T_{C}=29$ K for $p=216$ GPa \cite{Sakata}; the highest observed value of the critical temperature among all elements.

The above experimental results inspired us to estimate the basic thermodynamic parameters of the Ca superconducting state in the range of the very high pressures. In particular, we assume: $p=200$ GPa. The calculations will be conducted in the framework of the Eliashberg formalism, with the aid of the Eliashberg function calculated by Yin {\it et al.} in the paper \cite{Yin} ({\it Pnma} structure).

\section{THE ELIASHBERG EQUATIONS}

The Eliashberg equations have been derived in order to quantitative description of the superconducting state for the intermediate and strong coupling between electrons and phonons.

In appliance to the approach used in the BCS theory \cite{BCS}, the Eliashberg equations enable the detailed consideration of the complicated form of the electron-phonon interaction. In the result, the Eliashberg set allows exactly to estimate the value of the critical temperature, the order parameter and the electron effective mass, if the value of the Coulomb pseudopotential is known.

The Eliashberg equations can be written on the imaginary axis, on the real axis or in the mixed representation \cite{Eliashberg}. 
From the numerical point of view, the most convenient in application is the mixed representation, because the rather simple algorithms can be used \cite{SzczesniakA}, \cite{SzczesniakB}.

The Eliashberg equations in the mixed representation take the form \cite{Marsiglio}:
\begin{eqnarray}
\label{r1}
\phi\left(\omega\right)&=&
\frac{\pi}{\beta}\sum_{m=-M}^{M}\frac{\lambda\left(\omega-i\omega_{m}\right)-\mu^{*}\theta\left(\omega_{c}-|\omega_{m}|\right)}
{\sqrt{\omega_m^2Z^{2}_{m}+\phi^{2}_{m}}}\phi_{m}\\ \nonumber
                              &+& i\pi\int_{0}^{+\infty}d\omega^{'}\alpha^{2}F\left(\omega^{'}\right)
                                  [\left[N\left(\omega^{'}\right)+f\left(\omega^{'}-\omega\right)\right]\\ \nonumber
                              &\times&K\left(\omega,-\omega^{'}\right)\phi\left(\omega-\omega^{'}\right)]\\ \nonumber
                              &+& i\pi\int_{0}^{+\infty}d\omega^{'}\alpha^{2}F\left(\omega^{'}\right)
                                  [\left[N\left(\omega^{'}\right)+f\left(\omega^{'}+\omega\right)\right]\\ \nonumber
                              &\times&K\left(\omega,\omega^{'}\right)\phi\left(\omega+\omega^{'}\right)],
\end{eqnarray}
and
\begin{eqnarray}
\label{r2}
Z\left(\omega\right)&=&
                                  1+\frac{i\pi}{\omega\beta}\sum_{m=-M}^{M}
                                  \frac{\lambda\left(\omega-i\omega_{m}\right)\omega_{m}}{\sqrt{\omega_m^2Z^{2}_{m}+\phi^{2}_{m}}}Z_{m}\\ \nonumber
                              &+&\frac{i\pi}{\omega}\int_{0}^{+\infty}d\omega^{'}\alpha^{2}F\left(\omega^{'}\right)
                                  [\left[N\left(\omega^{'}\right)+f\left(\omega^{'}-\omega\right)\right]\\ \nonumber
                              &\times&K\left(\omega,-\omega^{'}\right)\left(\omega-\omega^{'}\right)Z\left(\omega-\omega^{'}\right)]\\ \nonumber
                              &+&\frac{i\pi}{\omega}\int_{0}^{+\infty}d\omega^{'}\alpha^{2}F\left(\omega^{'}\right)
                                  [\left[N\left(\omega^{'}\right)+f\left(\omega^{'}+\omega\right)\right]\\ \nonumber
                              &\times&K\left(\omega,\omega^{'}\right)\left(\omega+\omega^{'}\right)Z\left(\omega+\omega^{'}\right)], 
\end{eqnarray}
where: $K\left(\omega,\omega^{'}\right)\equiv \frac{1}{\sqrt{\left(\omega+\omega^{'}\right)^{2}Z^{2}\left(\omega+\omega^{'}\right)-\phi^{2}\left(\omega+\omega^{'}\right)}}$.

The symbols: $\phi\left(\omega\right)$, $Z\left(\omega\right)$, ($\phi_{n}\equiv\phi\left(i\omega_{n}\right)$ and 
$Z_{n}\equiv Z\left(i\omega_{n}\right)$) represent the order parameter function and the wave function renormalization factor on the real (imaginary) axis, respectively; $\omega_{n}$ is the $n$-th Matsubara frequency: $\omega_{n}\equiv \left(\pi / \beta\right)\left(2n-1\right)$. The inverse temperature is given by: $\beta\equiv\left(k_{B}T\right)^{-1}$, and $k_{B}$ is the Boltzmann constant. The order parameter is defined as: $\Delta\equiv \phi/Z$. 

%
\begin{figure}[t]%
\includegraphics*[scale=0.29]{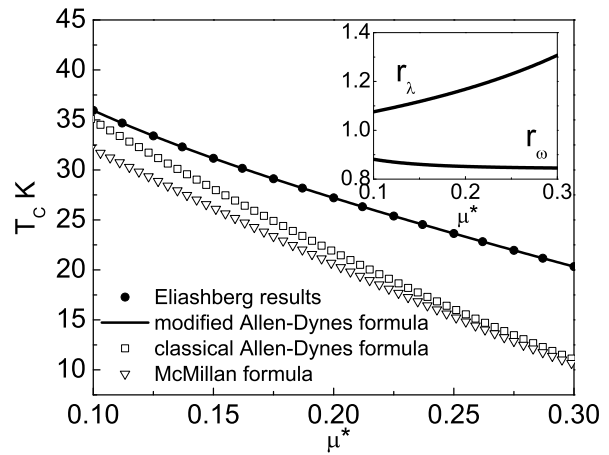}
\caption{The dependence of the critical temperature on the Coulomb pseudopotential. In the inset, we have shown the ratios $r_{\omega}$ and $r_{\lambda}$ as the function of the Coulomb pseudopotential.}
\label{f1} 
\end{figure}
%

In the framework of the Eliashberg formalism, the pairing kernel for the electron-phonon interaction is determined in the following way: $\lambda\left(z\right)\equiv 2\int_0^{\Omega_{\rm{max}}}d\Omega\frac{\Omega}{\Omega ^2-z^{2}}\alpha^{2}F\left(\Omega\right)$. The symbol $\alpha^{2}F\left(\Omega\right)$ denotes the Eliashberg function \cite{Yin}; the value of the maximum phonon frequency ($\Omega_{\rm{max}}$) is equal to $78.11$ meV. 

The depairing Coulomb interaction is modeled parametrically with the aid of the Coulomb pseudopotential $\mu^{*}$. Symbol $\theta$ denotes the Heaviside unit function and $\omega_{c}$ is the cut-off frequency ($\omega_{c}=3\Omega_{\rm{max}}$). The Bose-Einstein and Fermi-Dirac functions are represented by $N\left(\omega\right)$ and $f\left(\omega\right)$, respectively.

The Eliashberg equations have been solved numerically for $2201$ Matsubara frequencies ($M=1100$). The stability of the solutions has been achieved for the temperatures greater or equal to $T_{0}\equiv 8.12$ K. Additional information on the means of the analysis of the Eliashberg equations, the reader can find in the papers: \cite{Varelogiannis}, \cite{SzczesniakC}, and \cite{SzczesniakD}.

\section{THE RESULTS}

%
\begin{figure}[t]%
\includegraphics*[scale=0.29]{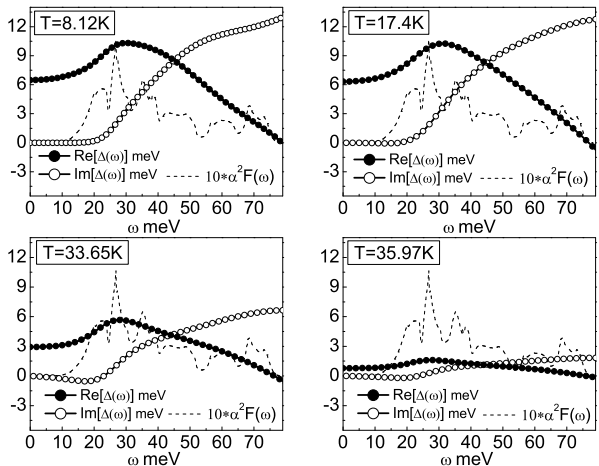}
\caption{The real and imaginary part of the order parameter on the real axis for selected values of the temperature. The rescaled Eliashberg function has been also plotted.} 
\label{f2}
\end{figure}
%
\begin{figure}[t]%
\includegraphics*[scale=0.29]{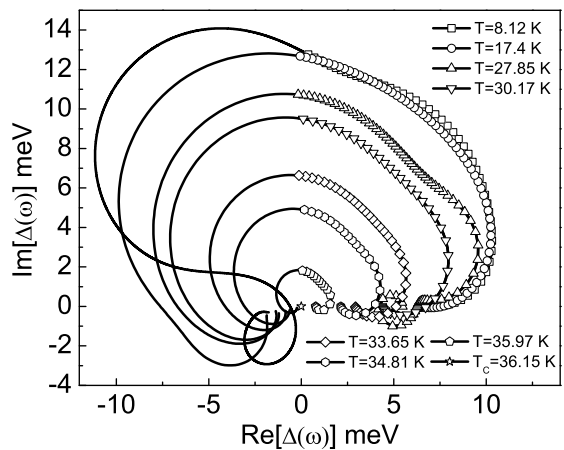}
\caption{The form of the order parameter on the complex plane for selected values of the temperature. The results obtained for $\omega\in\left<0,\Omega_{\rm max}\right>$ by using the lines with empty symbols have been denoted; the solid lines represent results achieved for $\omega\in\left(\Omega_{\rm max},\omega_{c}\right>$.} 
\label{f3}
\end{figure}

In the Eliashberg formalism, the critical temperature depends on the form of the Eliashberg function and the value of the Coulomb pseudopotential. In the opposition to the Eliashberg function, the parameter $\mu^{*}$ is really hard to calculate using the {\it ab initio} methods. According to the above, we have taken into consideration the very wide range of the Coulomb pseudopotential's values e.g. $\mu^{*}\in\left<0.1,0.3\right>$. The significant suggestion, that not only low values of $\mu^{*}$ should be considered, are the results obtained for Calcium under pressure at $120$ GPa, where $\mu^{*}=0.215$ has been noted \cite{SzczesniakD}.

In \fig{f1}, we have presented the critical temperature as the function of $\mu^{*}$. The exact results obtained by using the Eliashberg equations are represented by the filled circles. It can be easily noticed, that together with increase of the parameter $\mu^{*}$, the critical temperature decreases from the value of $36.15$ K to $20.79$ K. The above result proves, that even for the very large value of Coulomb pseudopotential, $T_{C}$ achieves high values. We notice, that the range of the calculated values of $T_{C}$ agrees qualitatively with the recently experimental data. In particular, Sakata {\it et al.} have shown that: $T_{C}=29$ K for $p=216$ GPa \cite{Sakata}.

From the mathematical point of view the determination of the critical temperature with the aid of the Eliashberg equations is the complicated and time consuming issue. For that reason, we have presented the analytical formula, which exactly reproduces the numerical results in opposition to the classical expressions of Allen-Dynes or McMillan (see \fig{f1}) \cite{AllenDynes}, \cite{McMillan}. In order to do that, the Allen-Dynes formula with the newly fitted parameters has been used. We notice, that the fitted parameters have been determined on the basis of the $250$ exact values of $T_{C}\left(\mu^{*}\right)$, and the least squares method. The result takes the form:
\begin{equation}
\label{r3}
k_{B}T_{C}=f_{1}f_{2}\frac{\omega_{\rm ln}}{1.45}\exp\left[\frac{-1.03\left(1+\lambda\right)}{\lambda-\mu^{*}\left(1+0.06\lambda\right)}\right],
\end{equation}
where the strong-coupling correction function ($f_{1}$) and the shape correction function ($f_{2}$) are given by the expressions: 
$f_{1}\equiv\left[1+\left(\frac{\lambda}{\Lambda_{1}}\right)^{\frac{3}{2}}\right]^{\frac{1}{3}}$ and 
$f_{2}\equiv 1+\frac{\left(\frac{\sqrt{\omega_{2}}}{\omega_{\rm{ln}}}-1\right)\lambda^{2}}{\lambda^{2}+\Lambda^{2}_{2}}$.
The symbol $\lambda$ denotes the electron-phonon coupling constant: 
$\lambda\equiv 2\int^{\Omega_{\rm{max}}}_{0}d\Omega\frac{\alpha^{2}F\left(\Omega\right)}{\Omega}$. 
The quantity $\omega_{2}$ represents the second moment of the normalized weight function: 
$\omega_{2}\equiv \frac{2}{\lambda}\int^{\Omega_{\rm{max}}}_{0}d\Omega\alpha^{2}F\left(\Omega\right)\Omega$, and 
$\omega_{{\rm ln}}$ is called the logarithmic phonon frequency: 
$\omega_{{\rm ln}}\equiv \exp\left[\frac{2}{\lambda}\int^{\Omega_{\rm{max}}}_{0}d\Omega\frac{\alpha^{2}F\left(\Omega\right)}
{\Omega}\ln\left(\Omega\right)\right]$. In particular, for Calcium under the pressure at $200$ GPa, the following values have been obtained: $\lambda=1.23$, $\sqrt{\omega_{2}}=29.98$ $\rm{meV}$, and $\omega_{{\rm ln}}=35.92$ meV. The functions $\Lambda_{1}$ and $\Lambda_{2}$ are defined as: $\Lambda_{1}\equiv 2.6\left(1+1.8\mu^{*}\right)$, and 
$\Lambda_{2}\equiv 0.092\left(1-150\mu^{*}\right)\left(\frac{\sqrt{\omega_{2}}}{\omega_{\rm{ln}}}\right)$.

The modified Allen-Dynes (mAD) formula possesses the fitted parameters, which are sharply different than the parameters included in the classical expression derived by Allen and Dynes (AD). The above result is connected with the fact, that the effective phonon frequency 
($\left[\omega_{\rm eff}\right]_{\rm AD}$) and the effective coupling constant ($\left[\lambda_{\rm eff}\right]_{\rm AD}$) in classical Allen-Dynes formula well reconstruct the exact value of $T_{C}$ only for low values of the Coulomb pseudopotential ($\mu^{*}\rightarrow 0.1$). We notice, that the parameters $\omega_{\rm eff}$ and $\lambda_{\rm eff}$ have been defined as: $k_{B}T_{C}=\omega_{\rm eff}e^{-\frac{1}{\lambda_{\rm eff}}}$. In the case of high values of $\mu^{*}$, the classical parameterization overestimates the effective phonon frequency and much underestimates the effective coupling constant. The inset in the Fig. 1 presents the detailed dependences of the ratios 
$r_{\omega}\equiv\left[\omega_{\rm eff}\right]_{\rm mAD}/\left[\omega_{\rm eff}\right]_{\rm AD}$ and 
$r_{\lambda}\equiv\left[\lambda_{\rm eff}\right]_{\rm mAD}/\left[\lambda_{\rm eff}\right]_{\rm AD}$ on $\mu^{*}$.

Finally, we notice that the classical expressions for $T_{C} $ should be applied with the great carefulness, because even for low-temperature superconductors like Al or Pb, the difference between analytical and numerical Eliashberg results are noticeable \cite{SzczesniakE}.


In \fig{f2}, we have presented the order parameter on the real axis for the selected values of the temperature and $\mu^{*}=0.1$. We see, that for the low frequencies, the non-zero value is taken only by the real part of $\Delta\left(\omega\right)$. The obtained result proves, that in the considered range of frequencies the damping effects not exist \cite{Varelogiannis}. For the higher values of the frequency ($\omega\sim 30$ meV) one can observe the maximum of Re[$\Delta\left(\omega\right)$], which is induced by the area of the characteristic peaks in the Eliashberg function. From the physical point of view, the above fact indicates that the relevant growth of Re[$\Delta\left(\omega\right)$] exists in the frequencies' area in which the electron-phonon coupling is exceptionally strong. Additionally, it should be marked out, that in the considered case the function Im[$\Delta\left(\omega\right)$] increases monotonically together with the growth of $\omega$. For the frequencies higher than $\sim 30$ meV, the real part of the order parameter decreases together with the growth of $\omega$ due to the drop of the Eliashberg function's value. In the last step let us turn our attention to the fact, that the growth of the Coulomb pseudopotential causes only the significant decrease of the order parameter while the shapes of Re[$\Delta\left(\omega\right)$] and Im[$\Delta\left(\omega\right)$] remain very similar.

In \fig{f3}, we have plotted the order parameter on the complex plane. It can be easily noticed, that the values of $\Delta\left(\omega\right)$ make the characteristic spirals with the radius that decreases together with the growth of the temperature. The obtained result allows to characterize the effective electron-electron interaction. In particular, for Re[$\Delta\left(\omega\right)$]$>0$ the effective interaction is pairing, whereas for Re[$\Delta\left(\omega\right)$]$<0$ the depairing processes are in major. In the case when $\mu^{*}=0.1$, the area of the frequencies in which the electron-electron interaction is attractive covers the range of the frequencies for which the Eliashberg function is definite. As the result of the $\mu^{*}$ increasing, the frequencies that correspond to the positive value of Re[$\Delta\left(\omega\right)$] become narrow from the side of $\Omega_{{\rm max}}$; however such effect is not exceptionally strong. 

%
\begin{figure}[t]%
\includegraphics*[scale=0.29]{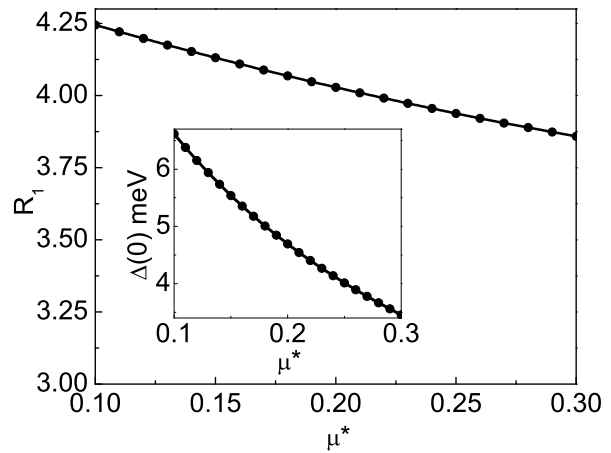}
\caption{The dependence of the ratio $R_{1}$ on the value of the Coulomb pseudopotential. In the inset, the influence of $\mu^{*}$ on the value of the parameter $\Delta\left(0\right)$ has been presented.} 
\label{f4}
\end{figure}
%
\begin{figure}[t]%
\includegraphics*[scale=0.29]{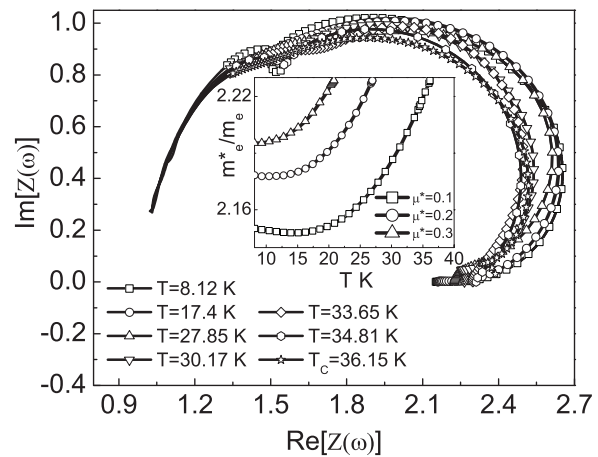}
\caption{The form of the wave function renormalization factor on the complex plane for selected values of the temperature and $\mu^{*}=0.1$. The results for $\omega\in\left<0,\Omega_{\rm max}\right>$ by using the lines with empty symbols have been denoted; the solid lines represent the results for $\omega\in\left(\Omega_{\rm max},\omega_{c}\right>$. In the inset, we have presented the dependence of the ratio $m^{*}_{e}/m_{e}$ on the temperature for selected values of the Coulomb pseudopotential.}
\label{f5}
\end{figure}
%

On the basis of the obtained results, we have calculated the physical value of the order parameter for the given temperature ($\Delta\left(T\right)$). In order to achieve that, one should use the following equation \cite{Eliashberg}: $\Delta\left(T\right)={\rm Re}\left[\Delta\left(\omega=\Delta\left(T\right)\right)\right]$. From the physical point of view, the most interesting is the value of the order parameter close to the temperature of zero Kelvin ($\Delta\left(0\right)\simeq\Delta\left(T_{0}\right)$), because it allows to determine the characteristic ratio: $R_{1}\equiv 2\Delta\left(0\right)/k_{B}T_{C}$. Let us notice, that in the framework of the BCS theory, the parameter $R_{1}$ takes the universal value equal to $3.53$ \cite{BCS}.

In \fig{f4}, we have presented the results achieved for Calcium, where the dependence of $R_{1}$ on the Coulomb pseudopotential has been plotted. We see, that the parameter $R_{1}$ is always bigger than $\left[R_{1}\right]_{\rm BCS}$ and only slightly decreases together with the increase of  $\mu^{*}$ (from $4.25$ to $3.90$). We notice, that the weak dependence of the ratio $R_{1}$ on $\mu^{*}$ is connected with the fact that the Coulomb pseudopotential in the comparable way lowers the value of $\Delta\left(0\right)$ and $T_{C}$. In particular, the open dependence of $\Delta\left(0\right)$ on $\mu^{*}$ in the Fig's. 4 inset have been plotted, whereas the shape of the function $T_{C}\left(\mu^{*}\right)$ in \fig{f1} has been shown.

From the physical point of view, the high values of the ratio $\left[R_{1}\right]_{\rm Ca}$ are the consequence of the existence of the strong-coupling and retardation effects, which are omitted by the simple BCS theory. In the framework of the Eliashberg approach, these effects can be characterized with the aid of the parameter $k_{B}T_{C}/\omega_{\rm ln}$. In the weak-coupling limit, one can assume: $\left[k_{B}T_{C}/\omega_{\rm ln}\right]_{\rm BCS}=0$. In the case of Calcium, we have obtained: $\left[k_{B}T_{C}/\omega_{\rm ln}\right]_{\mu^{*}=0.1}\simeq 0.104$, and      
$\left[k_{B}T_{C}/\omega_{\rm ln}\right]_{\mu^{*}=0.3}\simeq 0.059$. Above results clearly differ from the BCS approximation.

Finally, we notice that the exact value of $\mu^{*}$ for Calcium under the pressure at $200$ GPa is unknown. However, on the basis of the modified Allen-Dynes formula and the experimental result presented recently by Sakata {\it et al.} ($T_{C}=29$ K for $p=216$ GPa) \cite{Sakata}, one can obtain $\left[\mu^{*}\right]_{p=216 {\rm GPa}}\simeq  0.177$. If we assume that: $\left[\mu^{*}\right]_{p=200 {\rm GPa}}\sim\left[\mu^{*}\right]_{p=216 {\rm GPa}}$, the value of $\left[R_{1}\right]_{p=200 {\rm GPa}}$ equals about $4.0$.


In the Eliashberg formalism the dependence of the electron effective mass ($m^{*}_{e}$) on the temperature is determined by the wave function renormalization factor. In particular, we have: $m^{*}_{e}={\rm Re}\left[Z\left(0\right)\right]m_{e}$, where $m_{e}$ denotes the electron bare mass.

In \fig{f5}, we have plotted the values of $Z\left(\omega\right)$ on the complex plane in order to investigate the influence of the temperature on the form of the wave function renormalization factor. On the basis of the obtained results, it has been stated, that the wave function renormalization factor weakly depends on the temperature in comparison with the order parameter. We notice, that the increasing value of the Coulomb pseudopotential does not change significantly the course of $Z\left(\omega\right)$.

In the Fig's. 5 inset, the dependence of the ratio $m^{*}_{e}/m_{e}$ on the temperature for the selected values of the Coulomb pseudopotential has been shown. The achieved results prove, that the electron effective mass takes the very high value in the whole range of the superconducting phase's existence, and $m^{*}_{e}$ has the maximum for $T=T_{C}$, where $m^{*}_{e}/m_{e}=2.23$.
 
\section{Summary}

In the paper, we have obtained the basic thermodynamic parameters of the superconducting state in Calcium under the pressure at $200$ GPa. It has been stated, that dependently on the assumed value of the Coulomb pseudopotential ($\mu^{*}\in\left<0.1,0.3\right>$), the critical temperature decreases in the range from $36.15$ K to $20.79$ K. This result is in agreement with the recently experimental data obtained by Sakata {\it et al.} ( $T_{C}=29$ K for $p=216$ GPa) \cite{Sakata}.

Next, the dependence of the order parameter $\Delta\left(0\right)$ on $\mu^{*}$ has been determined. The results allowed to determine the value of the ratio $R_{1}$. We have affirmed, that the parameter $R_{1}$ is essentially larger than $\left[R_{1}\right]_{\rm BCS}$, and slightly decreases with the growth of the Coulomb pseudopotential from $4.25$ to $3.90$. 

In the last step, the electron effective mass has been determined. We have shown, that $m^{*}_{e}$ is large and takes its maximum for $T=T_{C}$. In particular: $\left[m^{*}_{e}\right]_{\rm max}=2.23 m_{e}$.

In the paper, the detailed characteristic of the order parameter on the real axis has been also presented. It has been proven, that the courses of Re[$\Delta\left(\omega\right)$] and Im[$\Delta\left(\omega\right)$] are clearly correlated with the shape of the Eliashberg function. Additionally, the plot of the order parameter on the complex plane enabled the conclusion, that for $\mu^{*}=0.1$, the effective electron-electron interaction is attractive in the frequencies' range, where the Eliashberg function is definite. The increase of $\mu^{*}$ causes only the insignificant narrowing of the frequencies' range from the side of $\Omega_{{\rm max}}$.

\begin{acknowledgments}
The authors would like to thank K. Dzili{\'{n}}ski for providing excellent working conditions and the financial support.

Some calculations have been conducted on the Cz{\c{e}}stochowa University of Technology cluster, built in the framework of the
PLATON project, no. POIG.02.03.00-00-028/08 - the service of the campus calculations U3.
\end{acknowledgments}


%
\end{document}